# Shedding new light on the Crab with polarized X-rays

M. Chauvin[1,2], H.-G. Florén[3], M. Friis[1,2], M. Jackson[1]†, T. Kamae[4,5], J. Kataoka[6], T. Kawano[7], M. Kiss[1,2], V. Mikhalev[1,2], T. Mizuno[7], N. Ohashi[7], T. Stana[1,2], H. Tajima[8], H. Takahashi[7], N. Uchida[7], M. Pearce[1,2]*

[1]KTH Royal Institute of Technology, Department of Physics, 106 91 Stockholm, Sweden.

[2]The Oskar Klein Centre for Cosmoparticle Physics, AlbaNova University Centre, 106 91 Stockholm, Sweden.

[3]Stockholm University, Department of Astronomy, 106 91 Stockholm, Sweden.

[4]University of Tokyo, Department of Physics, Tokyo 113-0033 Tokyo, Japan.

[5]SLAC/KIPAC, Stanford University, 2575 Sand Hill Road, Menlo Park, CA 94025, USA.

[6]Research Institute for Science and Engineering, Waseda University, Tokyo 169-8555, Japan.

[7]Hiroshima University, Department of Physical Science, Hiroshima 739-8526, Japan.

[8]Institute for Space-Earth Environment Research, Nagoya University, Aichi 464-8601, Japan.

*Correspondence to: pearce@kth.se.

†Now at School of Physics and Astronomy, Cardiff University, Cardiff CF24 3AA, UK.

Strong magnetic fields, synchrotron emission, and Compton scattering are omnipresent in compact celestial X-ray sources. Emissions in the X-ray energy band are consequently expected to be linearly polarized. X-ray polarimetry provides a unique diagnostic to study the location and fundamental mechanisms behind emission processes. The polarization of emissions from a bright celestial X-ray source, the Crab, is reported here for the first time in the hard X-ray band (~20-160 keV). The Crab is a complex system consisting of a central pulsar, a diffuse pulsar wind nebula, as well as structures in the inner nebula including a jet and torus. Measurements are made by a purpose-built and calibrated polarimeter, PoGO+. The polarization vector is found to be aligned with the spin axis of the pulsar for a polarization fraction, PF = (20.9 ± 5.0)%. This is higher than that of the optical diffuse nebula, implying a more compact emission site, though not as compact as, e.g., the synchrotron knot. Contrary to measurements at higher energies, no significant temporal evolution of phase-integrated polarisation parameters is observed. The polarization parameters for the pulsar itself are measured for the first time in the X-ray energy band and are consistent with observations at optical wavelengths.

## Introduction

The Crab is a prototypical celestial particle accelerator *(1).* The central pulsar comprises a highly magnetized (~$10^8$ T) neutron star. The rotation period, 33.7 ms, slows as $\dot{P} = 4.2 \times 10^{-13}$ *(2).* Of order 1% of the rotational energy loss is imparted to electrons (and positrons, here referred to as electrons) which can be accelerated up to an energy of several PeV ($10^{15}$ eV). Electrons are extracted along the boundary of the co-rotating magnetosphere and directed along open magnetic field lines to the light cylinder. The electrons pass through turbulent magnetic fields near and beyond the light cylinder and form an ultra-relativistic wind. As this wind expands into ejecta from the progenitor star and supernova explosion, wind termination shocks are formed. Resulting synchrotron and inverse Compton interactions generate the high luminosity (~$1.3 \times 10^{38}$ erg/s) arcminute-sized nebula *(1).* The pulsar wind nebula shows a wealth of smaller-scale structures which are known to be highly dynamic, varying on short time-scales, and in emission energy. Overall, the size of the emitting region decreases when observed at high energies, but X-ray emission is still observed even close to the boundary of the nebula *(3)*, evidence that the situation is complex. The observations presented here cover a large field-of-view, ~2°, encompassing the pulsar and entire nebula. The rotational phase of the pulsar is used to isolate the pulsar emissions from that of the nebula. The phase-folded light-curve as observed by PoGO+ *(4)* in the ~20-160 keV range is shown in Fig. 1. It is obtained by folding through a pulsar rotation period using the closest ephemeris to our observations from the Jodrell Bank Observatory *(5).* The peaks arise due to the offset between the rotation and magnetic axes of the pulsar. Two peaks are apparent - a main peak, P1, phase interval 0.06 - 0.14, and a second peak P2, phase interval 0.44 - 0.55. The phase intervals are defined in the same way as in *(6).* The pulsar light-curve is reproduced by a variety of models, e.g. *(7-10)*, where X-rays are generated in synchrotron emission from accelerated electrons. The polarization properties of the pulsar emission depend on the location of the emission region in the magnetosphere *(11)*. Emissions dominated by the nebula can be isolated by selecting X-rays from the off-pulse region, phase interval 0.64 - 1.0.

The optical polarization properties for a handful of rotation-powered pulsars have been determined. For the Crab, averaged over all phases, PF = (9.8 ± 0.1)% for the pulsar region *(12).* Phase-resolved polarimetry reveals rapid swings in the polarization angle, PA, and significantly reduced PF in both pulse peaks. This supports caustic emission from the outer magnetosphere *(9),* where the peak emission comprises radiation from a large range of altitudes (i.e. emission with different field directions, hence the swing in PA and destructive interference of PF), and/or emission from the equatorial current sheet where the magnetic field changes polarity *(10).* X-ray polarization is expected to track that observed at optical wavelengths since both emissions are synchrotron in nature, share the same magnetic field-lines and the same electron population *(11).* High-energy pulsar models are currently tested using optical polarization data. Until now, X-ray results stem solely from a polarimeter on-board OSO-8 *(13)*. Highly statistically significant results have been obtained for the nebula. A PF of (19.2 ± 1.0)% at a PA of (156.4 ± 1.4)° and (19.5 ± 2.8)% at a PA of (152.6 ± 4.0)° was determined at 2.6 and 5.2 keV, respectively. There is only marginal evidence of polarization for the pulsed part of the light curve *(14)*. At higher energies (>200 keV), polarimetric measurements have been reported by INTEGRAL, as described in the discussion below. The CZTI instrument on-board AstroSat *(15)* is expected to provide polarimetric data (>100 keV) there-by also complementing the results presented here.

**Results**

Results are presented in Fig. 1 and Fig. 2. Phase-integrated Crab emissions exhibit PF = (20.9 ± 5.0)%, providing a detection at more than 4σ significance. For synchrotron processes, the maximum allowed PF for a uniform magnetic field geometry is 60-75% *(16)*, for electron spectral indices in the range 1-3 *(6)*. Despite observations encompassing both the pulsar and the topologically complex wind nebula, a relatively high value of PF is found, indicating a magnetically ordered, and therefore compact, emission site. High-resolution X-ray images from Chandra *(17)* reveal a rich structure in the inner nebula. Two concentric magnetic tori are centered on the pulsar position. The inner torus lies in a plane perpendicular to the pulsar spin axis, whose projection onto the sky is (124.0 ± 0.1)° *(18)*. All angles are defined anticlockwise relative to North (i.e. to the East). Electrons trapped in the toroidal magnetic field produce synchrotron radiation with a PA parallel to the pulsar spin axis *(19)*. A PA value of (131.3 ± 6.8)° is determined, which coincides with that of the spin axis. This is in agreement with the expectation from NuSTAR imaging *(20)* showing the toroidal ring region dominating emission in the hard X-ray band of PoGO+. Optical polarization measurements have higher spatial resolution, which allows individual features to be discerned. Measurements with HST *(21)* find a high polarization fraction from the synchrotron knot (PF = 59.0 ± 1.9)% at a PA = (124.7 ± 1.0)°, as well as in the wisps at PAs of 124 - 130°. In contrast, a vector map of the entire inner nebula shows a peak distribution of PA around 165°. X-ray imaging cannot resolve such details, but the coincidence of the PA with observed structures points to these features being associated with the X-ray torus.

For PoGO+, the light-curve pulsed fraction contributes (18.5 ± 0.5)% to the total observed flux. Phase-integrated measurements are therefore indeed nebula-dominated, as conjectured above. The off-pulse region exhibits PF = ($17.4^{+8.6}_{-9.3}$)% and PA = (137 ± 15)°. The polarization properties of this nebula-dominated phase are compatible with the phase-integrated properties. Emissions from the pulsar peaks are isolated through phase selections and the constant off-pulse contribution from the nebula is subtracted. For P1, the 99% upper limit for PF is 70%. The PA is poorly constrained, see the Supplementary Information. For P2, PF = ($33.5^{+18.6}_{-22.3}$)% and PA = (86 ± 18)°.

**Discussion**

The SPI and IBIS instruments on board INTEGRAL have been used as polarimeters. It is important to note that neither instrument rotated during observations, and rely on Monte Carlo simulations to resolve angular dependencies in the instrument. This makes the determined polarization parameters vulnerable to systematic errors. Moreover, pre-launch polarimetric calibration has not been performed. The SPI team reports PF = (46 ± 10)% for PA = (123 ± 11)° for the off-pulse period of the Crab pulsar in the energy band 100 keV-1 MeV *(22)*. The IBIS team reports a phase-integrated result of PF = ($47^{+19}_{-13}$)% for PA = (100 ± 11)° for 200 - 800 keV *(23)*. More recent IBIS measurements reported a change in PA after a Crab flare event which may indicate the presence of magnetic field reconnection *(24)*. However, the measured PF is surprisingly large at > 60%.

Figure 3 summarizes existing off-pulse polarimetric observations of the Crab at high energies. Our off-pulse results show no significant changes in PF compared to that observed at optical wavelengths for the nebula. Considering the inconsistency of results between SPI and IBIS for similar energy ranges, our results favour the slower increase in PF with energy reported by SPI.

The P2 (off-pulse subtracted) measurements are the first in the hard X-ray regime. It is encouraging that similar behavior to optical observations is seen, as this regime is used to validate high-energy models, e.g. *(25)* found that only the polarization for the bridge emission between P1 and P2 was expected to change significantly with energy. In X-rays, the polarization parameters are integrated over the entire pulse which complicates interpretation. For comparison, measurements of P2 *(12)* with OPTIMA at the Nordic Optical Telescope, when integrated over the corresponding phase-region, yield PF = (6.85 ± 0.2)% and PA = (97.6 ± 0.2)°. The superior temporal resolution shows an angle swing through the peak of 100° (130° for P1).

Our off-pulse PA is also consistent with that reported in the optical regime and is parallel to the pulsar spin axis *(18)* as expected. This contrasts with the PA measured by OSO-8 which is 30-33° displaced from the spin axis. INTEGRAL instruments also reported off-pulse PAs consistent with the spin axis. Newer measurements *(24)* claim a >3σ difference with a phase-integrated PA of (80 ± 12)°. Off-pulse measurements have not been reported. Figure 4 also includes measurements from the PoGOLite Pathfinder instrument, a predecessor to PoGO+ *(26)*, performed in 2013. The measurements are 3 years apart (as compared to ~8 for IBIS). Although the detection of polarization by PoGOLite is marginal, no significant change is observed in either PF or PA for this energy interval.

**Summary**

A significant detection of polarized emission from the Crab system (including phase dependence) in the energy interval ~20-160 keV is reported for the first time. Measurements do not support the high PF value *(24)* derived using an X-ray instrument onboard INTEGRAL as a polarimeter. Moreover, no significant change in polarization angle is observed when comparing PoGO+ data to that obtained 3 years previously by PoGOLite. The PA for the Crab nebula is observed to be consistent with the projection of the pulsar spin axis in the plane of the sky, compatible with an origin in the toroidal magnetic field. It is tentatively confirmed that optical polarization data near the pulsar can be used as a proxy for higher energy emission when constructing pulsar models.

**Methods**

PoGO+ observations were conducted in July 2016 from a stabilized balloon-borne platform in the upper stratosphere (~40 km altitude) *(27)*. The polarimeter detection volume comprises an array of 61 plastic scintillator rods, each with hexagonal cross-section (12 cm long, ~3 cm wide). The 2° field-of-view of each rod is defined by a collimator. Polarized X-rays will Compton scatter preferentially in the direction perpendicular to the electric field vector *(28)*. A polarization event is defined by exactly 2 interactions in the scintillator array. Each event defines an azimuthal scattering angle in the plane of the sky. The distribution of angles is a harmonic function, the phase of which defines PA. The modulation amplitude defines PF. For an unpolarized beam, PF is Rayleigh-distributed and therefore positive definite. As a result, a large number of photons is required to make a statistically constrained measurement. Additionally, the design of the polarimeter must include a method to distinguish instrumental effects from source polarization. For PoGO+, this is achieved by rotating the polarimeter around the viewing axis during observations. This generates a continuous distribution of scattering angles and provides a uniform polarimetric response. The symmetric detector geometry and rotation allows the scattering angle distribution to be determined without the need for corrections from computer models.

The field-of-view is centered on the Crab with a precision better than 0.05° during observations. The effective area for polarization measurements is 3.8 $cm^2$ at 50 keV. Performance characteristics of the polarimter are detailed in the Supplementary Information. The polarization sensitivity ('modulation factor') for a 100% polarized beam is (37.8 ± 0.7)% *(4)*. Due to the positive definite nature of measurements, it is particularly important that the polarimeter response is determined using X-ray beams of known polarization, as well as unpolarized beams. The modulation factor for an unpolarized radiation source with an energy comparable to the median energy registered during Crab observations is (0.10 ± 0.12)%. Background impinging from outside the collimated field-of-view is mitigated with a segmented anticoincidence system and a passive polyethylene neutron shield. A residual background arises predominantly from neutrons scattered into the detection volume from the atmosphere *(29)*.

A total of 594419 polarization events were identified during 92 ks of Crab observations. The signal-to-background ratio is 0.142. Anisotropic background may cause a fake polarization signal. To address this, interspersed observations, totaling 79 ks, are conducted of fields 5° to the East and West of the Crab. Transition between Crab and background fields occurs every ~15 minutes, in order to track temporal behaviour. Polarization parameters are derived using unbinned and background-subtracted Stokes parameters *(30)*, as described in the Supplementary Information.

**Acknowledgments**

This research was supported in Sweden by The Swedish National Space Board, The Knut and Alice Wallenberg Foundation, and The Swedish Research Council. In Japan, support was provided by Japan Society for Promotion of Science and ISAS/JAXA. SSC are thanked for providing expert mission support and launch services at Esrange Space Centre. DST Control developed the PoGO+ attitude control system under the leadership of J.-E. Strömberg. Contributions from past Collaboration members and students are acknowledged. In particular, we thank M. Kole, E. Moretti, G. Olofsson and S. Rydström for their important contributions to the PoGOLite Pathfinder mission from which PoGO+ was developed.

**Author contributions**

M.C., H-G.F., M.F., M.J., T.Kam, J.K., T.Kaw, M.K., V.M., T.M., N.O., T.S., H.T., H.Tak., N.U. and M.P.contributed to the development of the PoGO+ mission concept and/or construction and testing of polarimeter hardware and software. Crab observations were conducted by M.C., H-G.F., M.F., M.K., V.M., T.S., H.Tak., N.U. and M.P.. Data reduction and analysis was performed by M.C., M.F., M.K., V.M., H.Tak. and M.P.. The manuscript was prepared by M.F., T.Kam., M.K., V.M. and M.P.. The mission principal investigator is M.P. and he lead the writing of the manuscript.

**Competing financial interests**

The authors declare no competing financial interests.

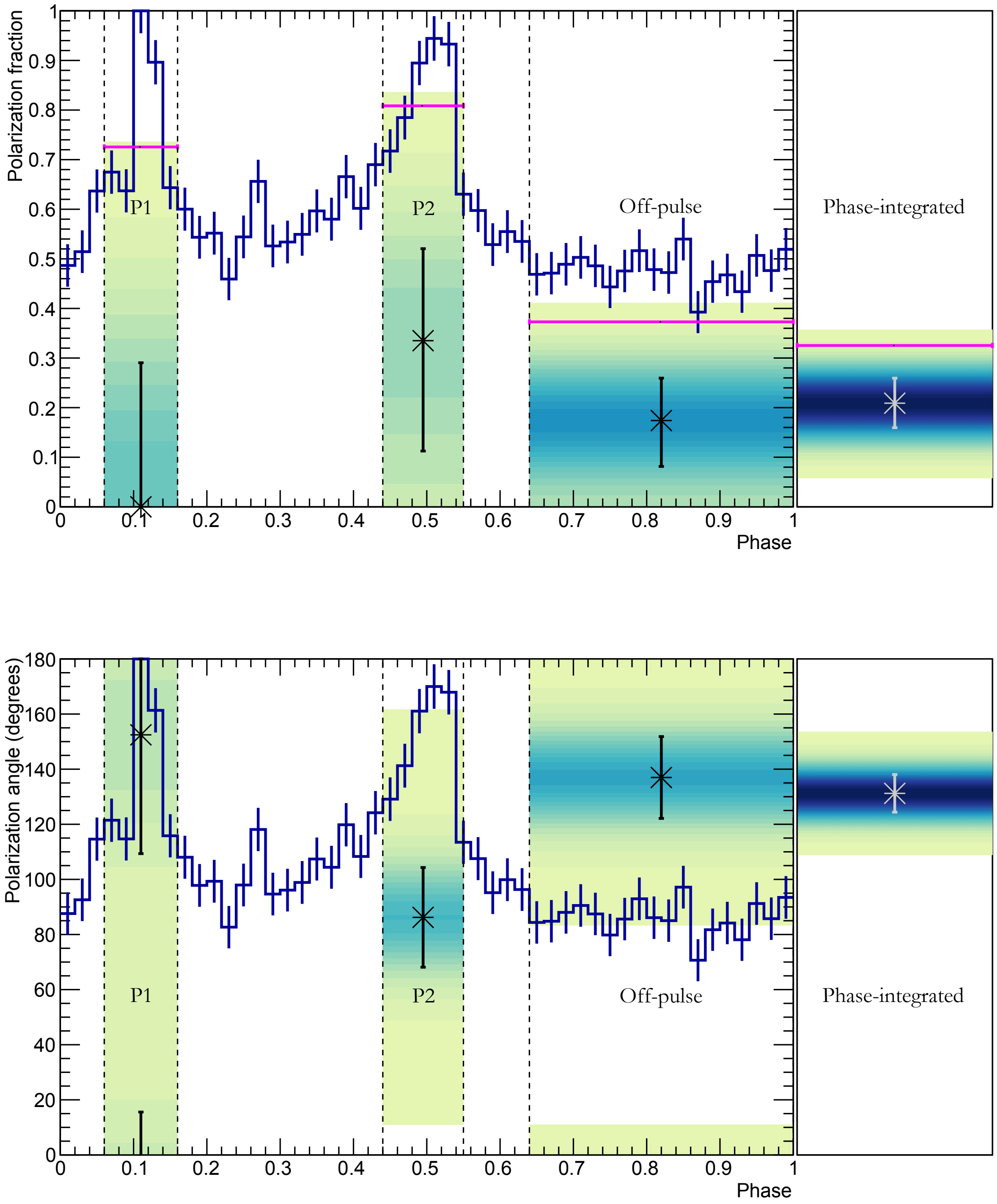

**Figure 1.** Light-curve and polarization results for the Crab.

Results for the polarization fraction (top) and polarization angle (bottom), super-imposed on the observed light-curve. The right-most column shows phase-integrated results. Colored overlays show the probability density distribution for the corresponding part of the light-curve (P1, P2, off-pulse, respectively). The off-pulse has been subtracted from P1 and P2 yielding a pure pulsar contribution. The error bars show the marginalized one standard-deviation Gaussian probability content while the magenta lines correspond to 99% upper limits (applicable to the polarization fraction only). The highest number of light-curve counts (corresponding to the peak of P1) is 2519.

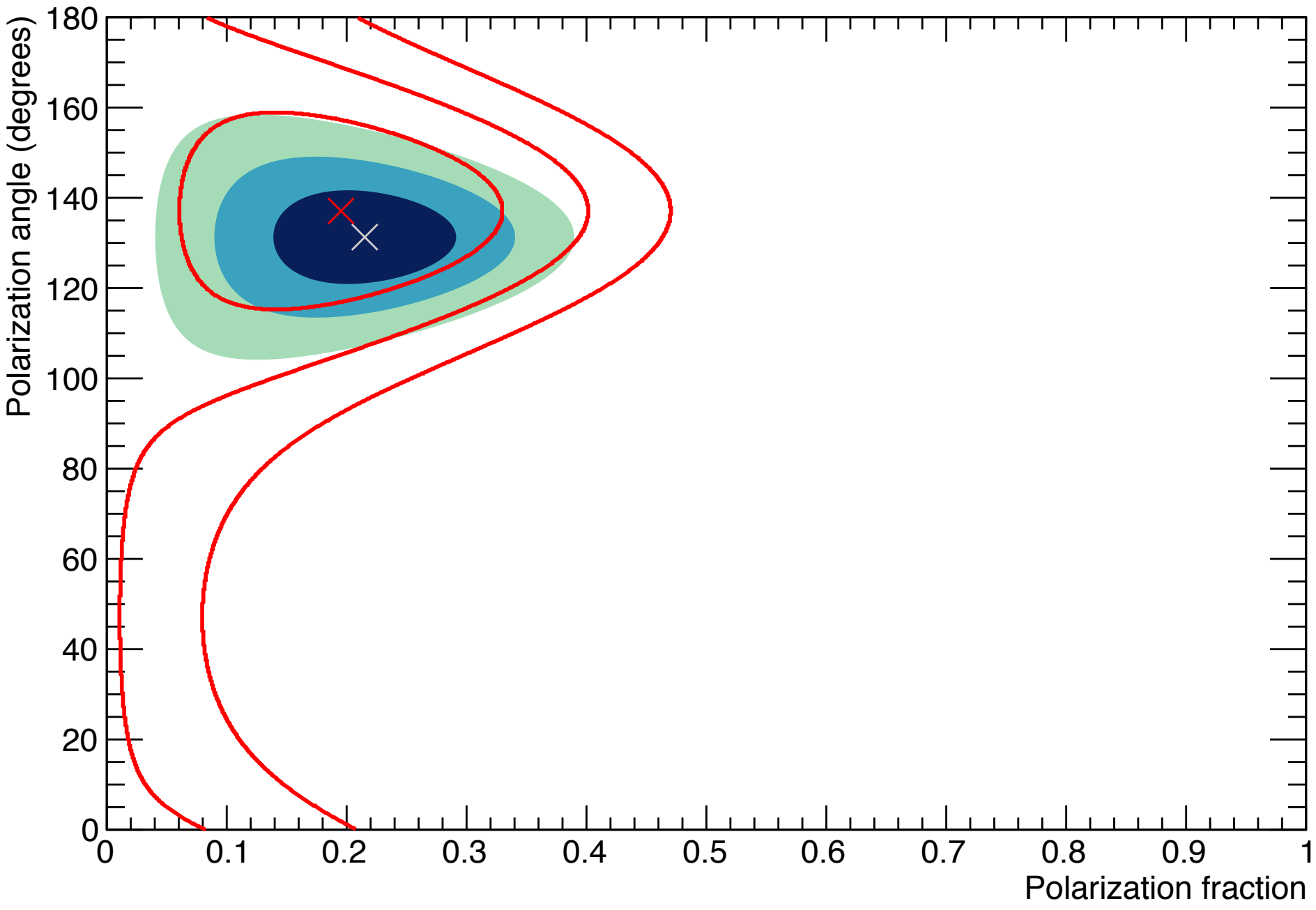

**Figure 2.** Contour plots for the Crab observation.

Gaussian 1, 2 and 3$\sigma$ probability contours for phase-integrated (shaded area) and off- pulse Crab observations (red lines). Crosses indicate maximum a posteriori estimates, see Supplementary Information for details.

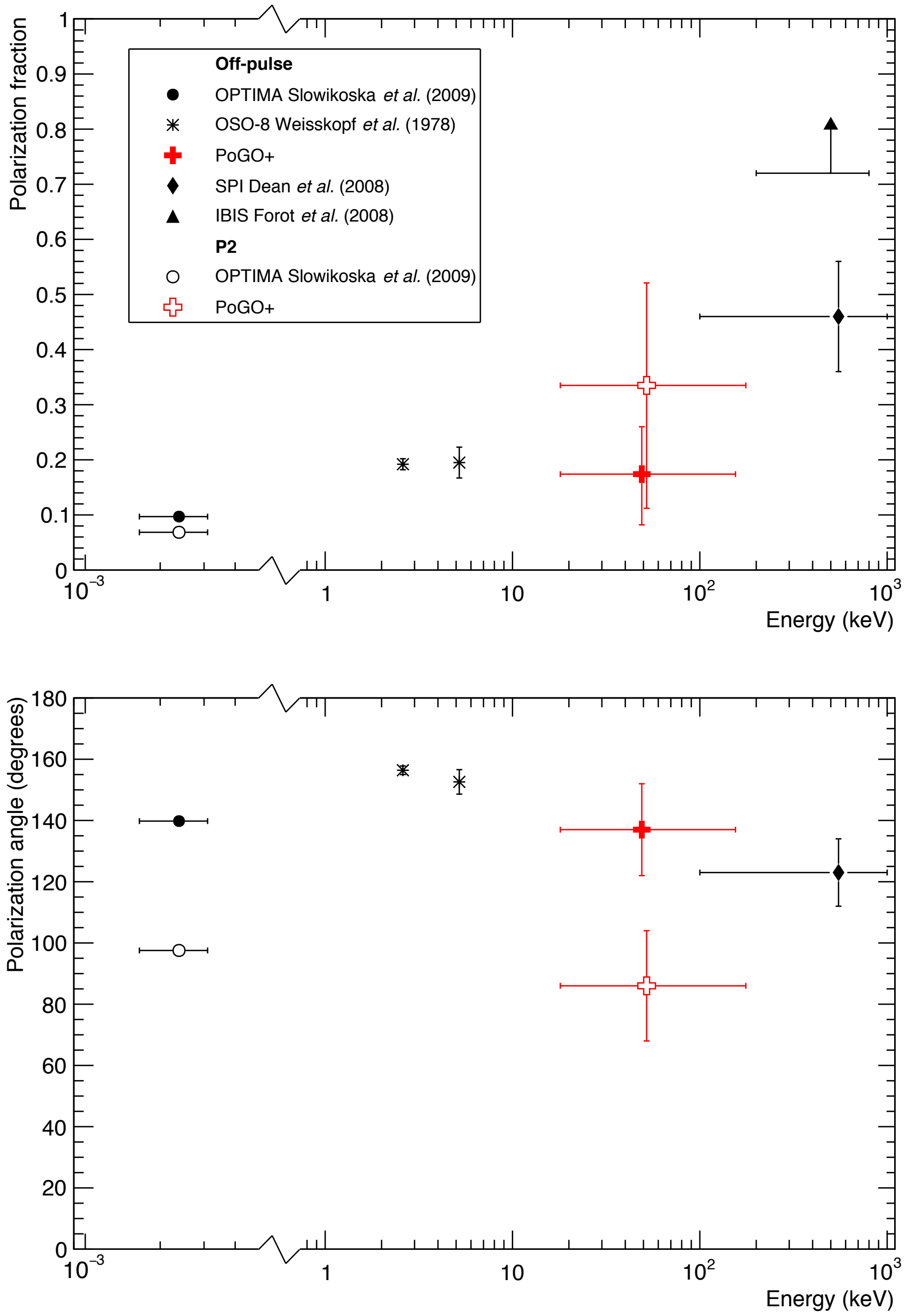

**Figure 3.** Comparison to other polarimetric studies of the Crab nebula (off-pulse) and P2. Data is shown for the PF (top row) and the PA (bottom row). It is noted that for optical results the nebula is spatially separated whereas temporal separation is applied in the X-ray regime.

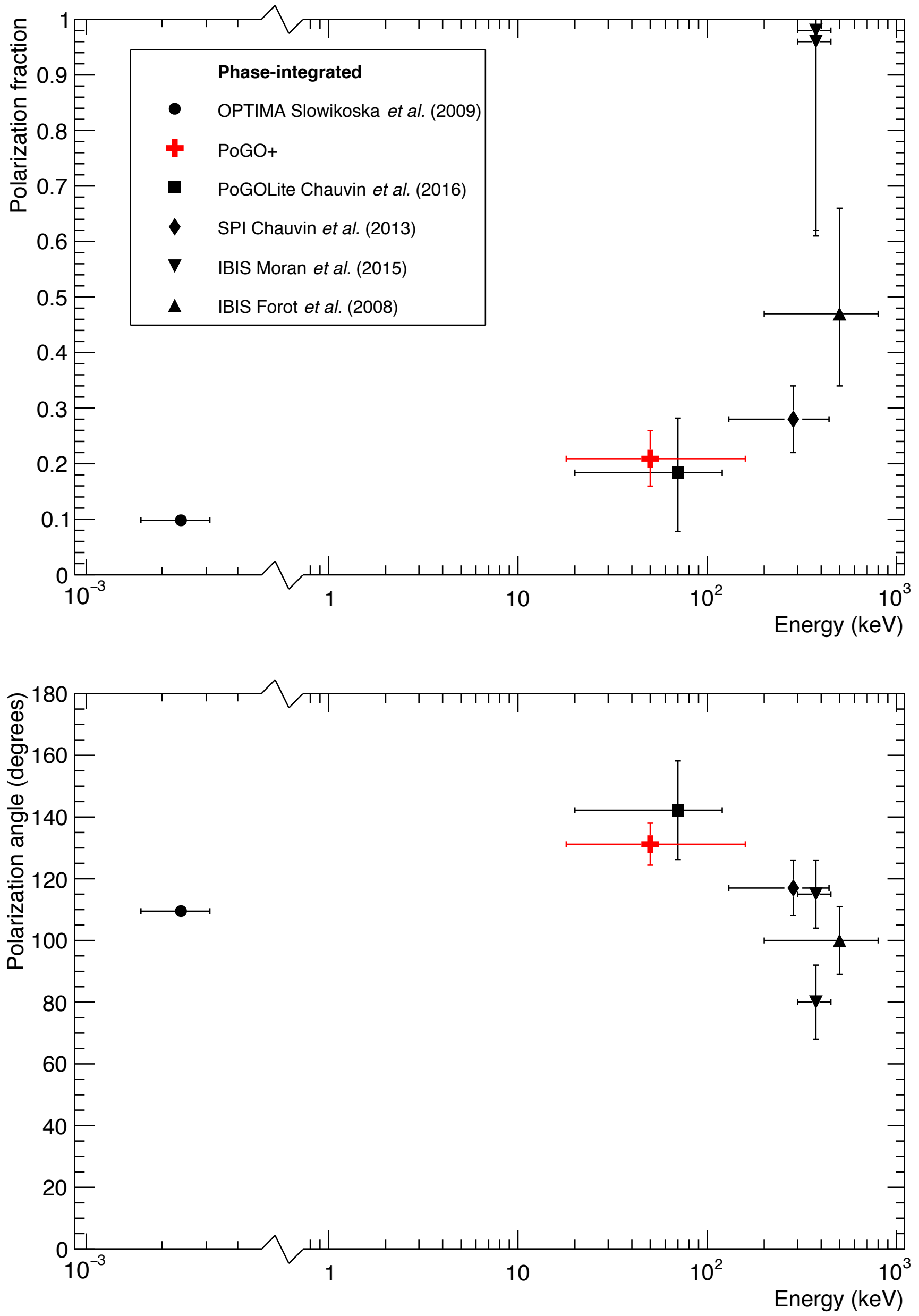

**Figure 4.** Comparison to other phase-integrated polarimetric studies of the Crab.

# *Supplementary information*

# Shedding new light on the Crab with polarized X-rays

M. Chauvin[1,2], H.-G. Florén[3], M. Friis[1,2], M. Jackson[1]†, T. Kamae[4,5], J. Kataoka[6], T. Kawano[7], M. Kiss[1,2], V. Mikhalev[1,2], T. Mizuno[7], N. Ohashi[7], T. Stana[1,2], H. Tajima[8], H. Takahashi[7], N. Uchida[7], M. Pearce[1,2]*

[1]KTH Royal Institute of Technology, Department of Physics, 106 91 Stockholm, Sweden.

[2]The Oskar Klein Centre for Cosmoparticle Physics, AlbaNova University Centre, 106 91 Stockholm, Sweden.

[3]Stockholm University, Department of Astronomy, 106 91 Stockholm, Sweden.

[4]University of Tokyo, Department of Physics, Tokyo 113-0033 Tokyo, Japan.

[5]SLAC/KIPAC, Stanford University, 2575 Sand Hill Road, Menlo Park, CA 94025, USA.

[6]Research Institute for Science and Engineering, Waseda University, Tokyo 169-8555, Japan.

[7]Hiroshima University, Department of Physical Science, Hiroshima 739-8526, Japan.

[8]Institute for Space-Earth Environment Research, Nagoya University, Aichi 464-8601, Japan.

*Correspondence to: pearce@kth.se.

†Now at School of Physics and Astronomy, Cardiff University, Cardiff CF24 3AA, UK.

**Definitions and theoretical framework**

The Klein-Nishina relationship describes the interaction of X-rays through Compton scattering. Photons will preferentially scatter in a direction perpendicular to their polarization vector. This yields a differential scattering cross-section

$$\frac{d\sigma}{d\Omega} \sim 2 \sin^2\theta \cos^2\phi$$

where, for a given range of polar scattering angles $\theta$, the azimuthal scattering angle $\phi$ will be modulated by the polarization of the incident beam *(28)*. The azimuthal scattering angle is derived from "2-hit" events (Compton-photoabsorption or Compton-Compton interactions) recorded in an array of detector pixels. For a polarized flux of photons, the resulting distribution of azimuthal scattering angles (or "modulation curve"), $f(\phi)$, exhibits harmonic behavior,

$$f(\phi) = N\{1 + A\cos[2(\phi - \phi_0)]\}$$

where $N$ is the number of 2-hit events, $A$ describes the ratio between the amplitude and mean value of the modulation, and $\phi_0$ is the phase of the modulation, related to the polarization angle as $\psi = \phi_0 - 90°$. The polarization signature has a 180° period, as indicated by the factor 2 in the argument of the cosine function. The value of $A$ and $\phi_0$ can be obtained through a chi-square fit to the modulation curve. For a measurement background with a flat distribution of azimuthal scattering angles, a modulation factor is defined as

$$M = A\left(1 + \frac{1}{R}\right)$$

where $R$ is the signal-to-background ratio for the observations. The polarization fraction is then given by

$$p = \frac{M}{M_{100}}$$

where $M_{100}$ is the modulation factor registered for a 100% polarized flux of photons. The value is defined by laboratory measurements and computer simulations and depends on the incident photon spectrum as well as on the properties of the detector system.

Inspection of the modulation curve provides an important quality control of measurement data allowing the presence of a possible background-induced lower order harmonic to be identified. In particular, anisotropic background incident on the polarimeter may introduce a 360° component, whereby the following modulation curve is needed to describe the data:

$$f(\phi) = N\{1 + A_{180}\cos[2(\phi_{180} - \phi_0)] + A_{360}\cos(\phi_{360} - \phi_0)\}$$

where subscripts "180" and "360" indicate parameters for the 180° and 360° harmonics, respectively.

To avoid binning systematics, a Stokes parameter-based approach can instead be adopted. Here, the Stokes parameters $Q$, $U$ and $I$ are defined as

$$Q = \sum_{i=1}^{N} w_i \cos(2\psi_i) \qquad U = \sum_{i=1}^{N} w_i \sin(2\psi_i) \qquad I = \sum_{i=1}^{N} w_i$$

where $N$ is the total number of measured photons. For a general treatment, a weight $w_i$ is additionally assigned to each photon such that

$$W^2 = \sum_{i=1}^{N} w_i^2$$

This weight can be used to account for different populations of events being over- or under-represented. There are several cases where this can occur: for polarimeters which are rotated in order to eliminate systematic effects (if all rotation angles are not equally populated), for events having different angular acceptance (polarimeters with asymmetric detector configurations such as square geometries) or for different exposure times for source and background measurements (live-time normalization).

Starting from formulas for weighted Stokes parameters *(30)*, the background-subtracted polarization fraction $p$ and polarization angle $\psi$ can be derived as

$$p = \frac{2}{M_{100}(I_{\mathrm{src+bg}} - I_{\mathrm{bg}})} \sqrt{\left(Q_{\mathrm{src+bg}} - Q_{\mathrm{bg}}\right)^2 + \left(U_{\mathrm{src+bg}} - U_{\mathrm{bg}}\right)^2}$$

$$\psi = \frac{1}{2} \tan^{-1}\left(\frac{U_{\mathrm{src+bg}} - U_{\mathrm{bg}}}{Q_{\mathrm{src+bg}} - Q_{\mathrm{bg}}}\right)$$

where "src+bg" and "bg" denote on-source and background observations, respectively.

The Minimum Detectable Polarization, MDP, is an established figure-of-merit for X-ray polarimeters. Based on *(30)*, it can be shown that for background-subtracted measurements at 99% confidence level, the quantity can be expressed as

$$\mathrm{MDP} = \frac{4.29\sqrt{W_{\mathrm{src+bg}}^{2} + W_{\mathrm{bg}}^{2}}}{M_{100}(I_{\mathrm{src+bg}} - I_{\mathrm{bg}})}$$

An unpolarized beam has a 1% probability of yielding a polarization fraction which exceeds the MDP due to statistical fluctuations. Only for measurements where $p \gg \mathrm{MDP}$ can $\sigma_p$ be considered Gaussian. In all other cases, due to the positive-definite nature of the measurement, $p$ will be biased, over-estimating the true polarization fraction $\Pi$. Furthermore, $\sigma_p$ will under-estimate the uncertainty. Due to angular symmetry, the polarization angle $\psi$ does not suffer from the same issue, although its uncertainty $\sigma_\psi$ will also be under-estimated, since it is inversely proportional to $\Pi$.

If the condition $p \gg \mathrm{MDP}$ is fulfilled, the uncertainties $\sigma_p$ and $\sigma_\psi$ can be calculated using standard error propagation. Neglecting the statistical uncertainty on the signal-to-background ratio and $M_{100}$ yields the following expressions:

$$\sigma_p = \frac{2}{M_{100}}\sqrt{\left(\frac{I_{\mathrm{src+bg}} + I_{\mathrm{bg}}}{2(I_{\mathrm{src+bg}} - I_{\mathrm{bg}})} - \frac{\Pi^2 {M_{100}}^2}{4}\right)\frac{{W_{\mathrm{src+bg}}}^2 + {W_{\mathrm{bg}}}^2}{{I_{\mathrm{src+bg}}}^2 - {I_{\mathrm{bg}}}^2}}$$

$$\sigma_{\psi} = \frac{1}{\Pi\sqrt{2\left(I_{\mathrm{src+bg}} - I_{\mathrm{bg}}\right)}}\sqrt{\frac{{W_{\mathrm{src+bg}}}^2 + {W_{\mathrm{bg}}}^2}{I_{\mathrm{src+bg}} - I_{\mathrm{bg}}}}$$

If the condition is not fulfilled, a more rigorous approach is required. The Central Limit Theorem dictates that *Q* and *U* are Gaussian distributed. Without loss of generality, it can be assumed that $\Psi = \frac{\pi}{8}$ as this removes the correlation between *Q* and *U*, resulting in the likelihood

$$\mathcal{L}(Q, U | Q_0, U_0) = \frac{1}{2\pi\sigma_Q\sigma_U}\exp\left[-\frac{1}{2}\left(\frac{(Q - Q_0)^2}{\sigma_Q^2} + \frac{(U - U_0)^2}{\sigma_U^2}\right)\right]$$

where $Q_0$ and $U_0$ are the true Stokes parameters. For compactness, the relation

$$\sigma \equiv \frac{2\sigma_Q}{M_{100}} = \frac{2\sigma_U}{M_{100}} = \frac{2}{M_{100}}\sqrt{\left(\frac{I_{\mathrm{src+bg}} + I_{\mathrm{bg}}}{2\left(I_{\mathrm{src+bg}} - I_{\mathrm{bg}}\right)} - \frac{\Pi^2 {M_{100}}^2}{8}\right)\frac{{W_{\mathrm{src+bg}}}^2 + {W_{\mathrm{bg}}}^2}{{I_{\mathrm{src+bg}}}^2 - {I_{\mathrm{bg}}}^2}}$$

is introduced. The likelihood for *p* and $\Delta\Psi = \psi - \Psi$, after a transformation from the Cartesian Stokes-system to a polar coordinate system, is then given by

$$\mathcal{L}\left(p, \Delta\Psi \,\middle|\, \Pi\right) = \frac{p}{\pi M_{100}\sigma^2}\exp\left(-\frac{p^2 + \Pi^2 - 2p\Pi\cos(2\Delta\Psi)}{2\sigma^2}\right)$$

which, for a uniform prior in polar coordinates *(30)*, yields the posterior

$$\text{Post}(\Pi, \Delta\Psi|p) = \frac{\mathcal{N}}{\sigma^2} \exp\left(-\frac{p^2 + \Pi^2 - 2p\Pi\cos(2\Delta\Psi)}{2\sigma^2}\right)$$

where $\mathcal{N}$ is the normalization. A good estimate for the maximum a posteriori probability (MAP) is $(\Pi = p, \Delta\Psi = 0)$, whereby the above equation simplifies to $\text{Post}(\Pi, \Delta\Psi|p) = \mathcal{N}/\sigma^2$. However, using $p$ as the point estimate of the posterior when marginalizing over the angle results in a bias such that $p > \Pi$. For a high measured polarization fraction with low uncertainty, the bias is modest, less than 1% relative, while for poorly constrained measurements close to or below MDP, it can be as high as 5% or more. This highlights the importance of using the marginalized posterior estimates in statistics-limited polarization measurements, especially when setting upper limits. Analogously, the uncertainties $\sigma_p$ and $\sigma_\psi$ are only proper for highly constrained polarization measurements, i.e. when $p \gg$ MDP. In all other cases, the uncertainties $\sigma_\Pi$ and $\sigma_\Psi$ on the marginalized posterior estimates of $\Pi$ and $\Psi$ should be used, which are given by the intervals of the highest probability density content corresponding to 68.7%.

**Design of the PoGO+ polarimeter**

PoGO+ is the successor to the balloon-borne PoGOLite Pathfinder *(27),* which flew in 2013 and provided a statistics-limited measurement of the phase-integrated Crab polarization *(26)*. The PoGO+ design, illustrated in Fig. S1, is developed based on experience from this flight, using detailed mass-modelling in the Geant4 framework *(32)*. Previously, tests at synchrotron facilities have been used to qualify the polarimeter concept, including the non-linear energy dependence of light yield in plastic scintillators

*(33)*. The PoGO+ simulation model is built bottom-up and validated step by step. Individual detector components are first characterized in laboratory measurements and results are compared to simulations. The simulation is then built around these components and results from the complete geometry are compared to calibration measurements with the assembled polarimeter detector array *(4)*. The bulk of polarimeter calibration tests with radioactive sources were conducted at the Esrange Space Centre contemporaneous with launch preparations in 2016.

A close-packed array of 61 optically isolated hexagonal cross-section plastic scintillator rods (EJ-204, 12 cm long, ~3 cm wide) is used to determine the azimuthal scattering angle of incident X-rays. Two plastic scintillator interactions within a coincidence window of 110 ns constitute a candidate “2-hit” polarization event, expected to be either a Compton scattering followed by a photoelectric absorption or two consecutive Compton scatterings. Each event can be time-tagged relative to Universal Time with a precision of ~100 ns. A lower anticoincidence shield is formed by 4 cm tall BGO ($Bi_4Ge_3O_{12}$) crystals glued to the base of each plastic scintillator, thereby also providing an interface between the hexagonal cross-section plastic scintillators and the round-windowed photomultiplier tubes (PMTs). The plastic and BGO scintillation light emissions have different decay times (1.8 ns and 300 ns, respectively) and pulse shape discrimination techniques are applied to distinguish the two contributions. The combination of a plastic scintillator and a bottom BGO element comprises a scintillator detector cell (SDC).

A hexagonal cross-section copper tube collimator (0.5 mm wall thickness, 67.5 cm length) is mounted in front of each scintillator rod, limiting the instrument field-of-view to ~2°. Each collimator is covered in foils of lead (100 μm) for increased stopping power and tin (100 μm) for mitigating fluorescence X-rays. The plastic scintillator rods are surrounded by a 30-unit side anticoincidence shield (SAS) made of BGO. Each SAS assembly is 60 cm tall, with a thickness between 3 cm (sides) and 4 cm (corners), and is read out by the same type of photomultipliers as used for the SDCs. The SDCs thus sit in the bottom of a 60 cm tall anticoincidence well with at least 3 cm thick BGO shields on each side, providing efficient protection from charged particle and photon backgrounds.

During observations, the polarimeter assembly is rotated back-and-forth around the viewing axis at a rate of 1°/s, which allows variations in detection efficiency between scintillator units to be addressed. For each polarization event, the rotation angle is obtained with a precision of ~0.01°. The rotation results in a smooth distribution of reconstructed scattering angles instead of the discretized curve that would otherwise arise when assuming center-to-center trajectory between hit detector units.

The flight background is dominated by the interaction of atmospheric neutrons, as demonstrated by Geant4 simulations *(29)*, validated through a dedicated balloon flight with neutron detectors *(34)*. Neutrons entering the plastic scintillator may undergo elastic scattering off a proton and fake a Compton scattering event. To mitigate this background, the anticoincidence shield is surrounded by a stationary polyethylene neutron shield which is 15 cm thick at the location of the plastic scintillator array. This shield reduces the background rate by an order of magnitude. Two LiCAF-based ($LiCaAlF_6$) neutron detectors *(34)* are mounted in the vicinity of the scintillator array for background monitoring during flight. These allow the rate of neutrons penetrating the polyethylene shield to be studied and also provides directional information, as the two units are placed on opposite sides of the detector array.

The data acquisition system *(4)* continuously samples all PMT waveforms at 100 MHz rate with 12 bit precision. The typical flight background rate is several hundred kHz, compared to the ~1 Hz 2-hit rate from the Crab. An on-line veto requiring that stored data meets several selection criteria reduces the read-out rate to ~3 kHz: (i) only events with 2 or more hits within a 110 ns coincidence window are stored (i.e. single-hit events, which are the most common interactions but carry no polarization information, are discarded), (ii) at least one hit in an event must exceed the trigger threshold of 5 keV to suppress low-energy background, (iii) an upper discriminator threshold, set at 86 keV, reduces the background from minimum ionizing cosmic ray particles and (iv), waveforms clearly originating from BGO are rejected based on their sampled pulse-shapes. A conservatively set “hit” threshold is additionally implemented for zero-suppression, discarding signals arising from electronics noise. Further background reduction is achieved through post-flight selections, as described below.

**Polarimetric performance**

The polarimetric response of PoGO+ has been determined using both polarized and unpolarized X-rays in the laboratory prior to flight *(4)*. These measurements use an Americium-241 radioactive source with the main emission line at 59.5 keV. By scattering the source photons through a 90° angle, a beam with ~100% polarization fraction and energy 53.3 keV is produced. The 59.5 keV and 53.3 keV energies are well-matched to the median energy of X-rays registered from the Crab (51 keV). For the unpolarized beam, a polarization fraction of $(0.10 \pm 0.12)\%$ was measured, which indicates that no fake polarization signal is induced for photons entering through the aperture. The absence of angular modulation for an unpolarized flux observed with PoGO+ is a direct result of the symmetric detector geometry and rotation of the polarimeter, as asymmetric and/or non-rotating segmented detector systems always exhibit some level of modulation, which must then be accounted for through simulations.

The effective area of the polarimeter is given in Fig. S2. It is based on *(4)* with modifications corresponding to flight conditions: (i) the trigger threshold is lowered from 10 keV to 5 keV, (ii) the Geant4 simulation uses the atmospheric column density profile as measured in flight instead of an assumed average column density and (iii) an additional normalization is applied to yield the same total count rate in the simulation as observed during the flight. This normalization factor is determined by simulating the Crab rate for the pulsar and nebula, shown in Fig. S3. The input spectrum for the nebula is a power law with the spectral index of -2.1. For the pulsar, the input spectrum uses the average spectral index observed by NuSTAR *(20)* weighted based on the shape of the light-curve (nebula-to-pulsar ratio) observed by PoGO+. The pulsed fraction of the light-curve (Fig. 1) constitutes $(18.5 \pm 0.5)\%$ of the total Crab rate, resulting in a pulsar spectral index of -1.9.

The parameter $M_{100}$ is determined as in *(4)* by applying flight selections to the ground calibration data. Comparing values for $M_{100}$ obtained through measurement in the laboratory and simulations there-of yields a scaling factor $S$ to be used between measurements and simulations. This scaling factor $S = M_{\mathrm{lab}}/M_{\mathrm{sim}}$ accounts for subtle differences such as mechanical offsets in the detector array and simplified geometries in the simulation. Although optimized independently and for different data-sets, there is no difference in scaling factor for ground calibration data when using "flight selections" and "ground calibration selections": $0.943 \pm 0.016$ and $0.945 \pm 0.019$, respectively. This demonstrates that the scaling factor *S* is not sensitive to the event selection, whereby the derived value can be used also for flight data to obtain properly scaled polarization results from the simulated values of $M_{100}$. Since $M_{100}$ depends on energy, different values are obtained for the Crab nebula, pulsar and combined flux, respectively, as presented in Table S1. This table also shows the corresponding energy ranges, defined as the limits outside which the count rate drops to less than 5% of the peak value. It is noted that despite the difference in spectral index between the pulsar and nebula, the difference in corresponding values for $M_{100}$ is small compared to the statistical precision of the measurement.

**Balloon flight**

Observations are conducted in the stratosphere at an altitude of ~40 km, corresponding to a zenith atmospheric overburden of ~4 g/cm$^2$. The constant contribution from ~0.1 g/cm$^2$ of passive material (e.g. scintillator wrappings and polarimeter window) in the polarimeter line-of-sight is negligible. Fig. S4 shows the gondola, which houses the polarimeter. It is suspended ~100 m below a 1.1 million cubic meter helium-filled balloon which has a diameter of ~150 m when fully developed. Neither the balloon nor gondola obstructs the polarimeter field-of-view during observations. The launch mass of the gondola is 1728 kg, with an additional 450 kg of ballast which is used to regain altitude for day-time Crab observations after the balloon gas has cooled during the night.

An attitude control system *(27)* ensures that the viewing axis of the polarimeter follows the sidereal motion of the Crab and compensates for perturbations such as torsional forces in the balloon rigging and stratospheric winds. The elevation axis of the polarimeter is mounted in a gimbal, with several motors used for in-flight attitude control. One motor controls the polarimeter elevation axle, while a motor connected to the balloon rigging (coarse regulation) and a motor-driven flywheel system (fine regulation) provide azimuthal positioning.

Motor control signals are formed using information from a number of independent attitude sensors. The attitude control loop operates at 100 Hz. During observations, the sun is separated by ~30° in azimuth from the Crab. A sun sensor is used as the primary attitude sensor augmented by an Inertial Measurement Unit (IMU) which corrects for high frequency disturbances coming from the balloon rigging, friction and polarimeter cabling. The intrinsic azimuth regulation precision is better than 0.01°. The elevation of the polarimeter is determined to better than 0.01° with respect to the gimbal structure by a rotary encoder. Corrections to the pointing direction and the polarimeter rotation angle are made based on the pitch and roll of the gondola with respect to the gravity vector of the Earth. The collimator geometry requires a modest pointing precision of order 0.1° so as not degrade the polarimetric performance due to detector shadowing effects. The pointing accuracy for Crab observations is independently determined using a daytime star camera which monitors the position of a nearby reference star, HIP 26451.

PoGO+ was launched from the Esrange Space Centre (68.89° N, 21.11° E) at 03:17 UT on July 12th 2016. The flight trajectory followed a line of approximately constant latitude from Esrange. The gondola was cut from the balloon at 21:38 UT on July 18th and landed on Victoria Island, Canada. Throughout the flight, the pointing precision (Gaussian standard deviation) was found to be well within the 0.1° design requirement.

**Observations and data reduction**

The Crab was observed on each day of the flight. The average overburden for observations, accounting for the elevation of the polarimeter, is 5.8 $g/cm^2$, with a corresponding transmission probability of 1%, 30% and 41% for 20 keV, 50 keV and 100 keV, respectively.

Each Crab observation was divided into 6 minute intervals corresponding to a 360° revolution of the polarimeter around the viewing axis. Off-source observations displaced 5° to the East and to the West of the Crab allow the polarization of the background to be determined. Pointing is asynchronous to the data acquisition. Throughout the flight, the following pointing scheme was repeated: Crab × 2, background West × 2, Crab × 2, background East × 2. Approximately equal exposure was thereby achieved for the Crab and for the background fields (~16 minutes on source, ~14 minutes off-source). As discussed in *(30)*, this is optimal for background-dominated observations. The interspersed background observations compensate for changes in source and background count rates, resulting from temporally changing behavior such as payload altitude, source elevation, temperatures, etc. The total observation time was 92 ks on the Crab and 79 ks for the background fields, where the dead-time of the data acquisition system has been accounted for. In order to minimize the MDP, only events recorded while the instrument pointing is within 0.5° of the Crab are accepted. Beyond this angular separation, more than 25% of the source flux is lost, resulting in a degradation of the signal-to-background ratio.

Selections developed during laboratory tests of the polarimeter *(4)* were applied to flight data. This is done in three steps: firstly, pre-trigger sample points are used to assess the quality of the sampled PMT waveform baseline. While large amplitude signals from cosmic ray events are discarded by the upper discriminator, they may cause the waveform baseline to be offset from the nominal level. By rejecting waveforms where the baseline is rising or falling steeply, the efficiency of the following pulse-shape discrimination techniques can be improved. Secondly, the coincidence of recorded detector hits is re-evaluated. Off-line, with unconstrained computational power compared to the on-board computer system, more sophisticated algorithms are used for determining the starting point of the waveforms. The coincidence window can thereby be reduced to 10 ns (as compared to 110 ns on-line), which results in a lower hit multiplicity for some events. Thirdly, pulse-shape discrimination is applied to the remaining waveforms as described in *(4)*, albeit with parameters re-optimized to yield the smallest MDP based on the total number of selected events and the signal-to-background ratio. This is achieved by applying identical selections to the on-source and the off-source data and then normalizing to live-time. Finally, a threshold of 8 g/cm$^2$ is applied to the column density, which is the choice that minimizes the MDP. These selection metrics are independent from polarization observables, i.e. do not introduce any bias to the data. The 2-hit count rate for on-source and off-source measurements during the flight is presented in Fig. S5. The MDP for the Crab observation, before background subtraction, is 10.1%, for a signal-to-background ratio of 0.1418 ± 0.0004. This comprises 594419 2-hit events from the Crab observations and 445199 2-hit events from the background fields. The final light-curve, Fig. 1, covers the energy range 19 – 160 keV.

**Polarization analysis**

Before the unbinned Stokes analysis is performed, the azimuthal angle distributions (2° binning) are inspected for $\chi^2/$ ndf goodness-of-fit, with respect to modulation curves with 180° and 360° components. The Gaussian uncertainty for each bin is calculated from the unweighted histogram, before correcting for over-represented polarimeter rotation angles (the data acquisition begins before the rotation motor is powered to ensure full 360° coverage). This procedure is applied to both the on-source and the off-source measurements. The background can be subtracted by taking the difference between these two histograms, where bin uncertainties are added in quadrature. Resulting distributions and fits for the on-source, off-source and background-subtracted on-source data are presented in Fig. S6, Fig. S7 and Fig. S8, respectively. Since the instrument response to an unpolarized beam has been demonstrated in laboratory measurements to be uniform (no modulation), the presence of a 360° component in the observations warrants an extended discussion, as presented in the Supplementary Text below.

Performing the Stokes analysis for the various light-curve phase cuts results in the MAP estimate $(p, \psi)$ together with the uncertainties $(\sigma_p, \sigma_\psi)$ as defined above. This approach is valid, since the statistical uncertainties on signal-to-background and $M_{100}$ are low (0.3% relative and 2% relative, respectively). The resulting parameters $(p, \psi)$ are then used to calculate the posterior distribution $\mathrm{Post}(\Pi, \Delta\Psi|p)$, shown in Fig. 2 for phase-integrated and off-pulse phase selections. The posterior is then marginalized to yield independent estimates of $\Pi$ and $\Psi$, as well as their uncertainties $\sigma_\Pi$ and $\sigma_\Psi$ and the 99% upper limit on $\Pi$. These results are summarized in Table S2. It is the marginalized estimates of $\Pi$ and $\Psi$ that are used for the discussions in the main text, not the MAP estimate $(p, \psi)$.

**Background anisotropy**

A prominent modulation (both 180° and 360° components) is seen in the off-source observation (Fig. S7). The segmentation of the side anticoincidence shield allows the anisotropy of the background flux to be studied independently from the polarization analysis. While events with interactions in the anticoincidence system are discarded by the on-line veto, pulse-height spectra are stored every minute independent of the global trigger from the data acquisition system. The energy range considered can be varied by integrating over different parts of the individual pulse-height spectra, e.g. including only saturated events (cosmic ray background) or only non-saturated events (low-energy background).

The relative count-rates in the side anticoincidence units are used to determine the anisotropy of the background flux incident on the polarimeter. Fig. S9 shows an example from a six-minute measurement, where each data point has contributions from six different detector units, spatially separated by 60° and temporally separated by one minute relative to each other. Data points have been fitted by the same function as used for the polarization analysis, having both a 180° and a 360° component. The phase of the 180° component is horizontal, while the 360° component is coming from the bottom of the instrument, i.e. scattering upwards (albedo background). Both phases are thus in agreement with results for the polarization measurement of the background (Fig. S7).

The LiCAF neutron detectors are used to determine the anisotropy in the neutron flux. A Stokes-based analysis is performed, where for each recorded neutron event, the position of the neutron detector relative to the zenith direction is used. The analysis yields a 180° and a 360° component, both of which align with the horizontal and vertical (albedo) directions as seen in the side anticoincidence system data, as well as in the polarization measurements of the background.

Table S3 shows combined results for all Crab on-source and off-source observations. The 180° and 360° components have the same behavior both in the background polarization (SDC units), in SAS intensity maps, and in the neutron scintillators. This indicates that the anisotropy observed during background measurements is incident on the polarimeter from the sides, i.e. not coming from within the aperture.

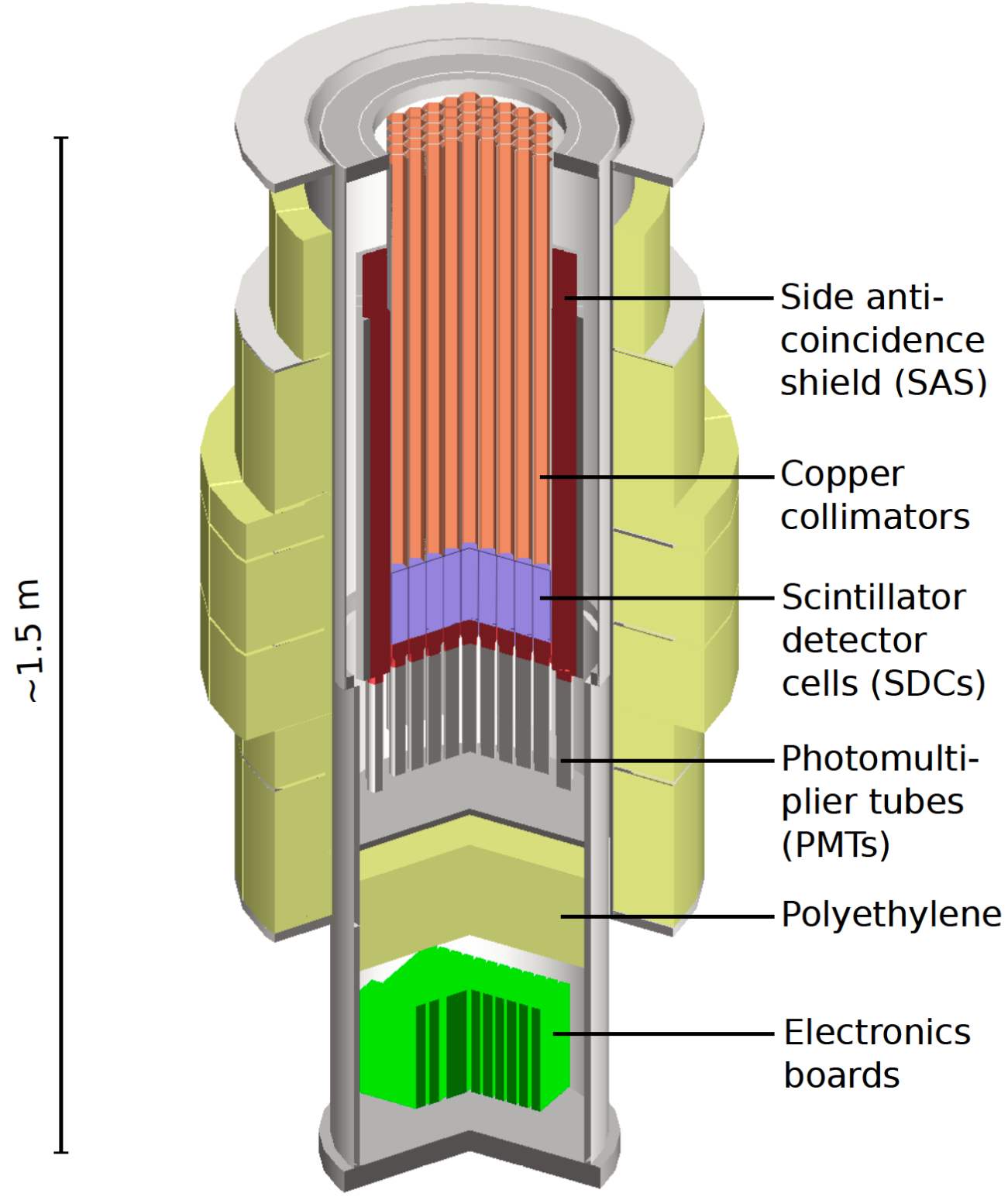

**Fig. S1. Design of the PoGO+ polarimeter.**

The scintillator detector array has an aperture diameter of approximately 30 cm. It is well-shielded by an active anticoincidence system as well as blocks of polyethylene for mitigating the neutron background.

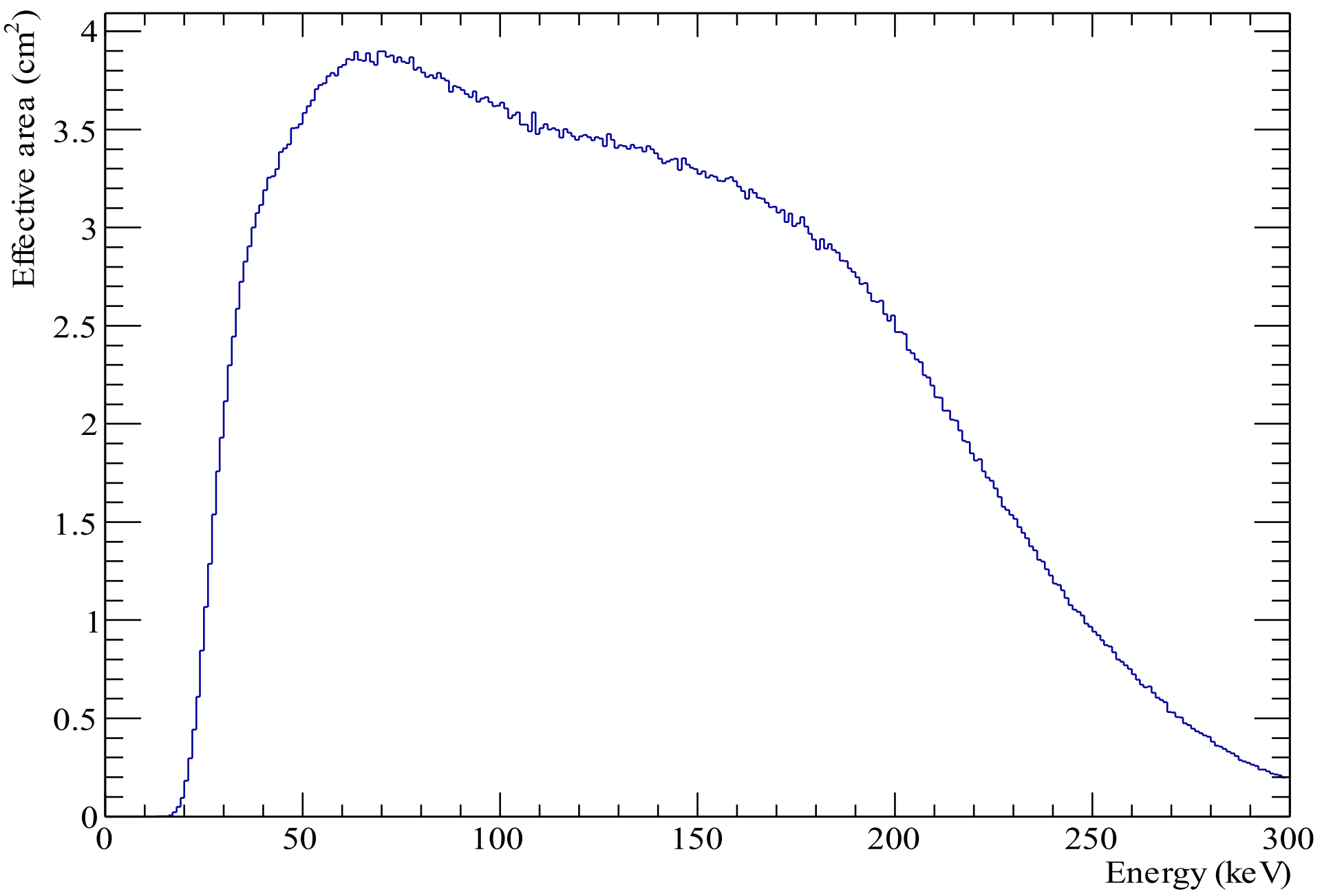

**Fig. S2. Simulated effective area as a function of energy.**

Events approximately in the range 20 – 160 keV are included for the polarization analysis of the Crab nebula data while for the pulsar, which has a harder spectrum, events up to about 180 keV are included.

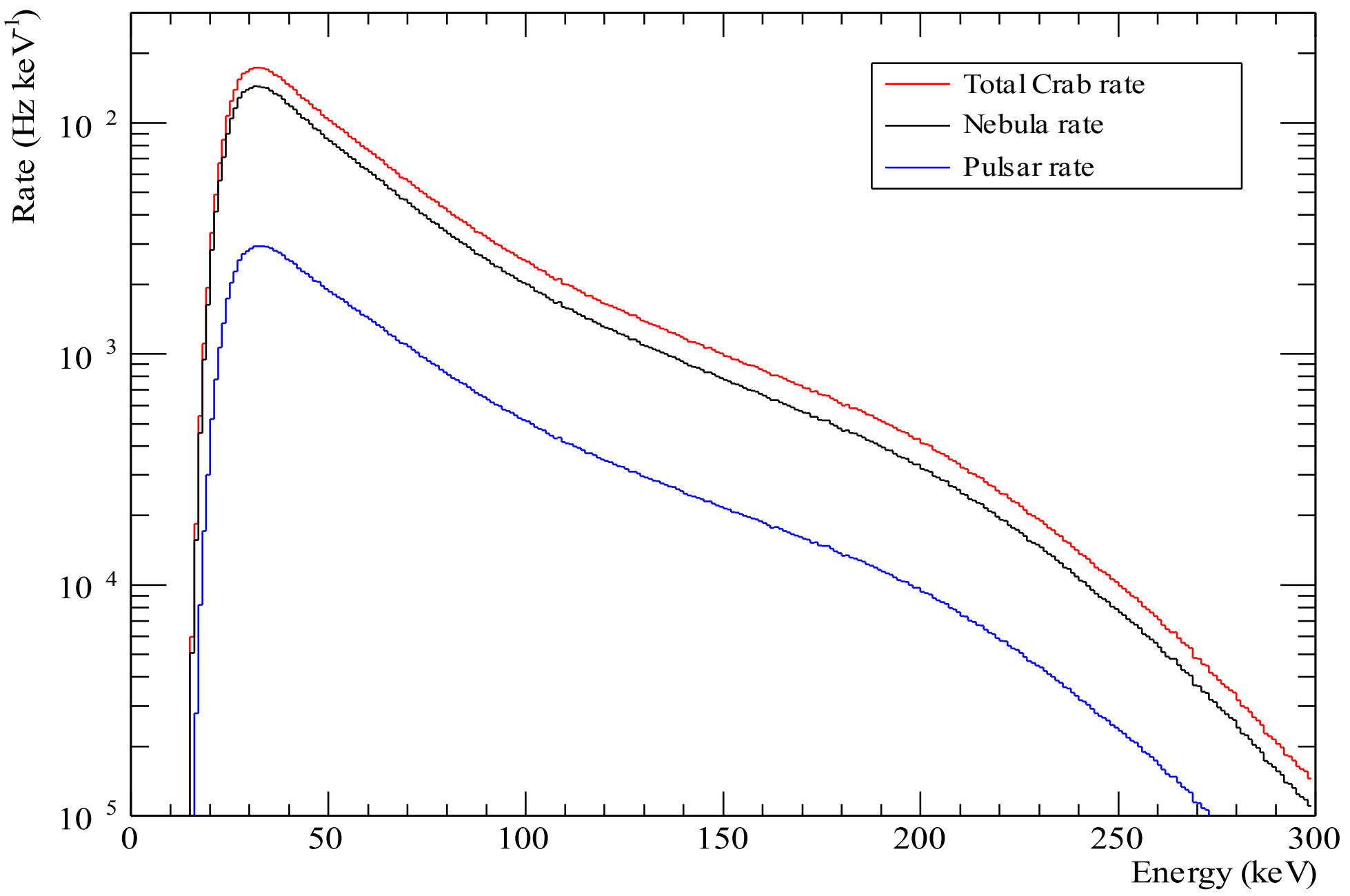

**Fig. S3. Simulated count rates for the total Crab, nebula and pulsar.**

Results are shown for the total Crab contribution in red (upper curve), nebula in black (middle curve) and pulsar in blue (bottom curve). The precise pulsar-to-nebula fraction is determined from the flight data. From this result, the energy range is defined as the limits (lower and upper) outside which the count rate drops to less than 5% of the peak value.

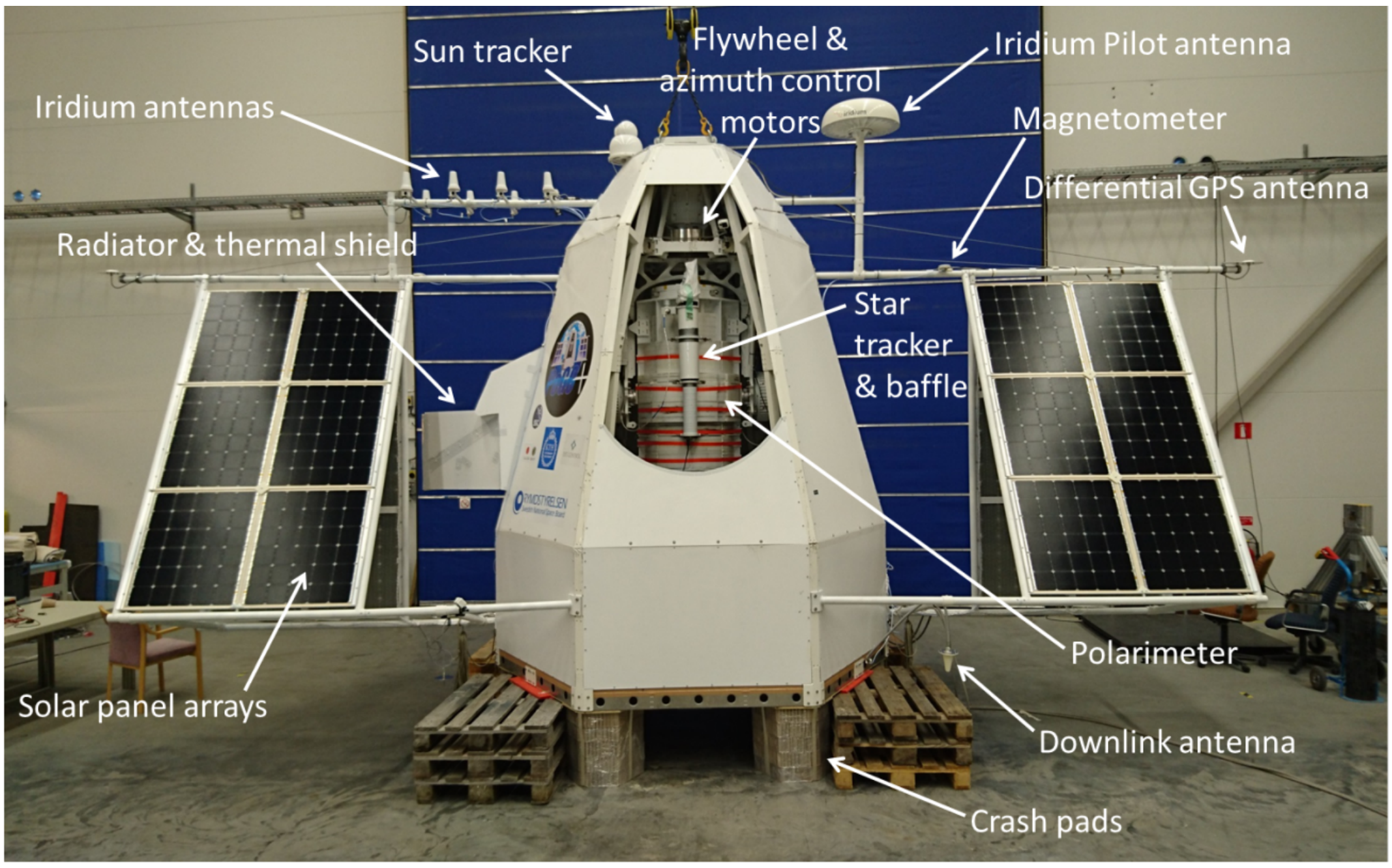

**Fig. S4. The PoGO+ payload.**

Gondola structure with solar panels, crash pads, communication antennas, etc. The polarimeter is stowed in the vertical position, whereby the co-axial star tracker camera and baffle assembly can be clearly seen.

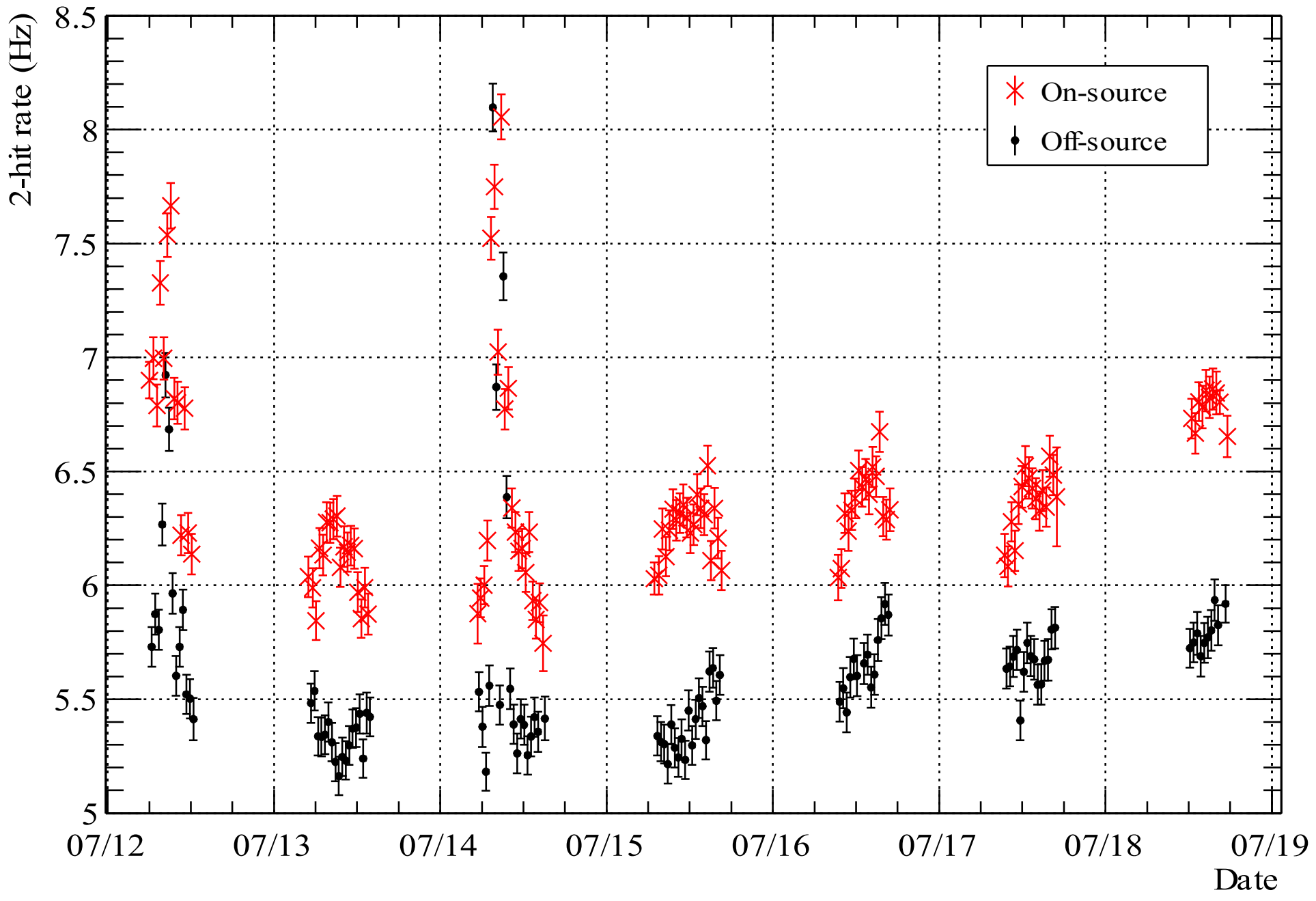

**Fig. S5. Count rate of 2-hit events as a function of time.**

Results are shown for on-source (red crosses) and off-source (black dots) observations as a function of date during the 2016 flight. Rapid changes in the rate are observed during the first day and third day of the flight. The changes are seen both in the on-source and off-source measurements, i.e. it is the signal-to-background ratio which is changing, not the source rate. These transients further underline the importance of recording interspersed background.

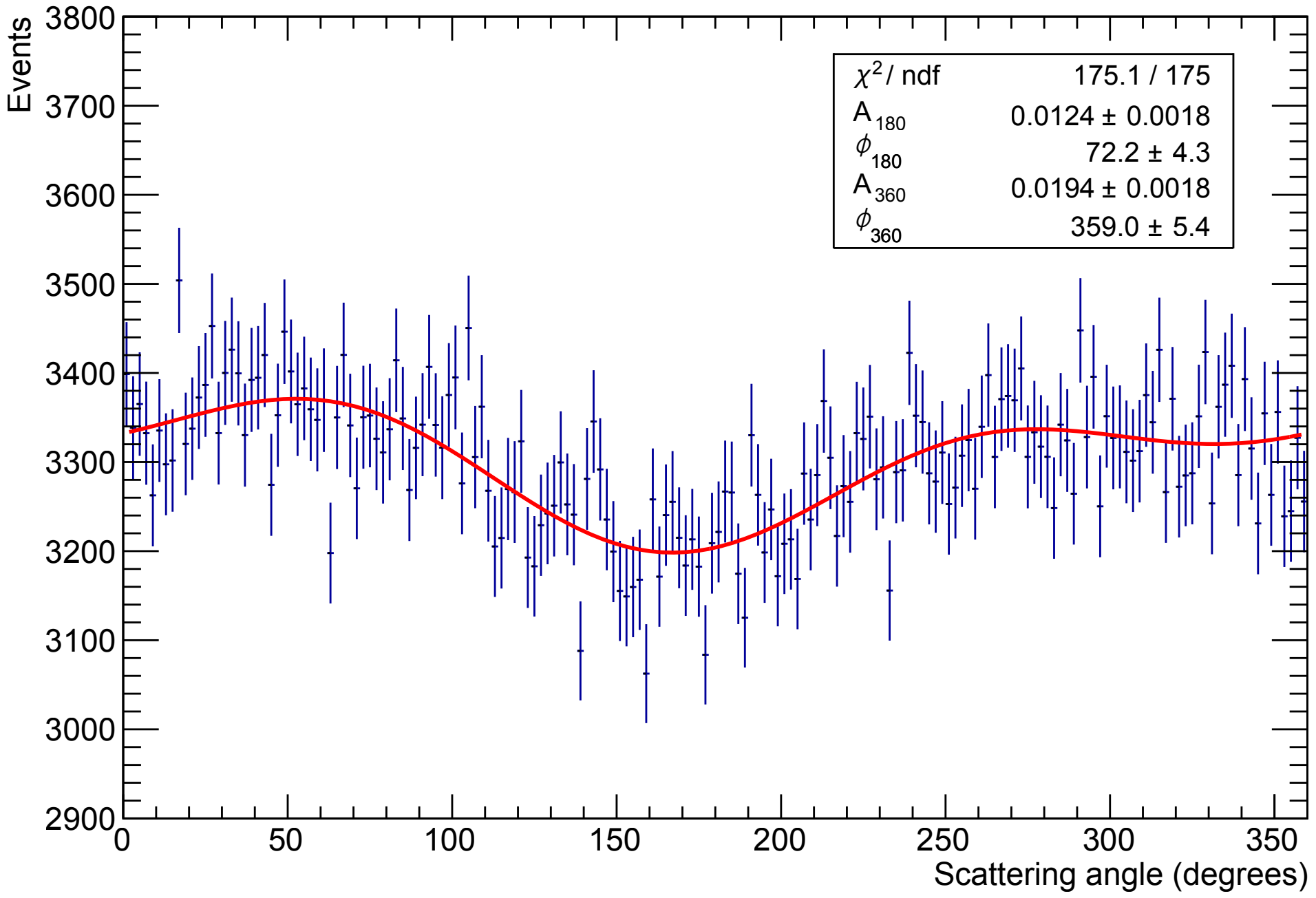

**Fig. S6. Modulation curve for on-source measurements.**

Scattering angle distribution and fitted modulation curve for the Crab observations. Both a 180° and a 360° component is needed to provide a good fit, indicating that the background incident on the polarimeter is anisotropic.

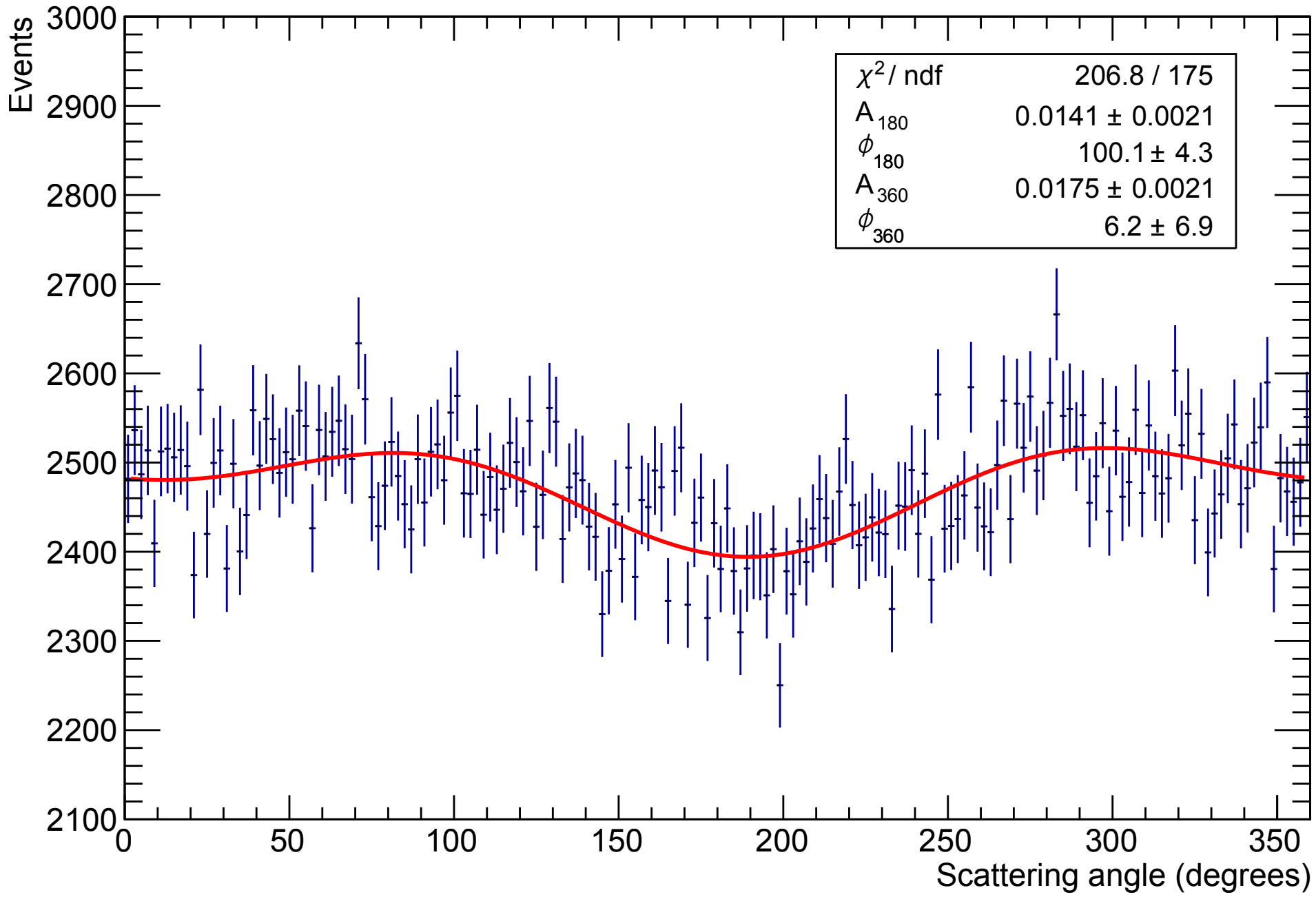

**Fig. S7. Modulation curve for off-source measurements.**

Scattering angle distribution and fitted modulation curve for the Crab background (off-source) observations. Both a 180° and a 360° component is needed to provide a good fit.

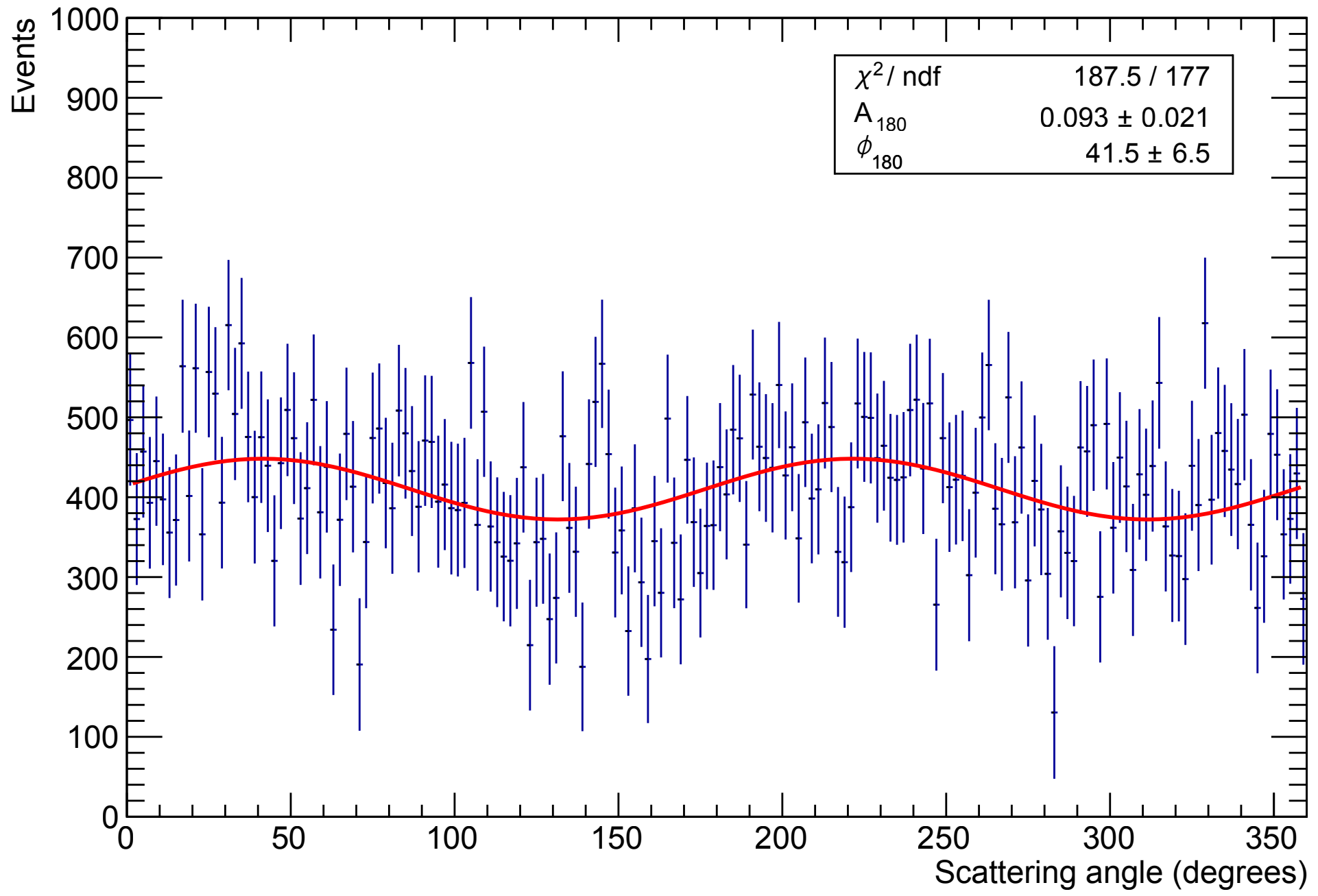

**Fig. S8. Modulation curve for the background-subtracted measurements.**

Scattering angle distribution and fitted modulation curve for the Crab observation following background subtraction. As indicated by the chi-square value, a 360° component is no longer needed to provide a good fit.

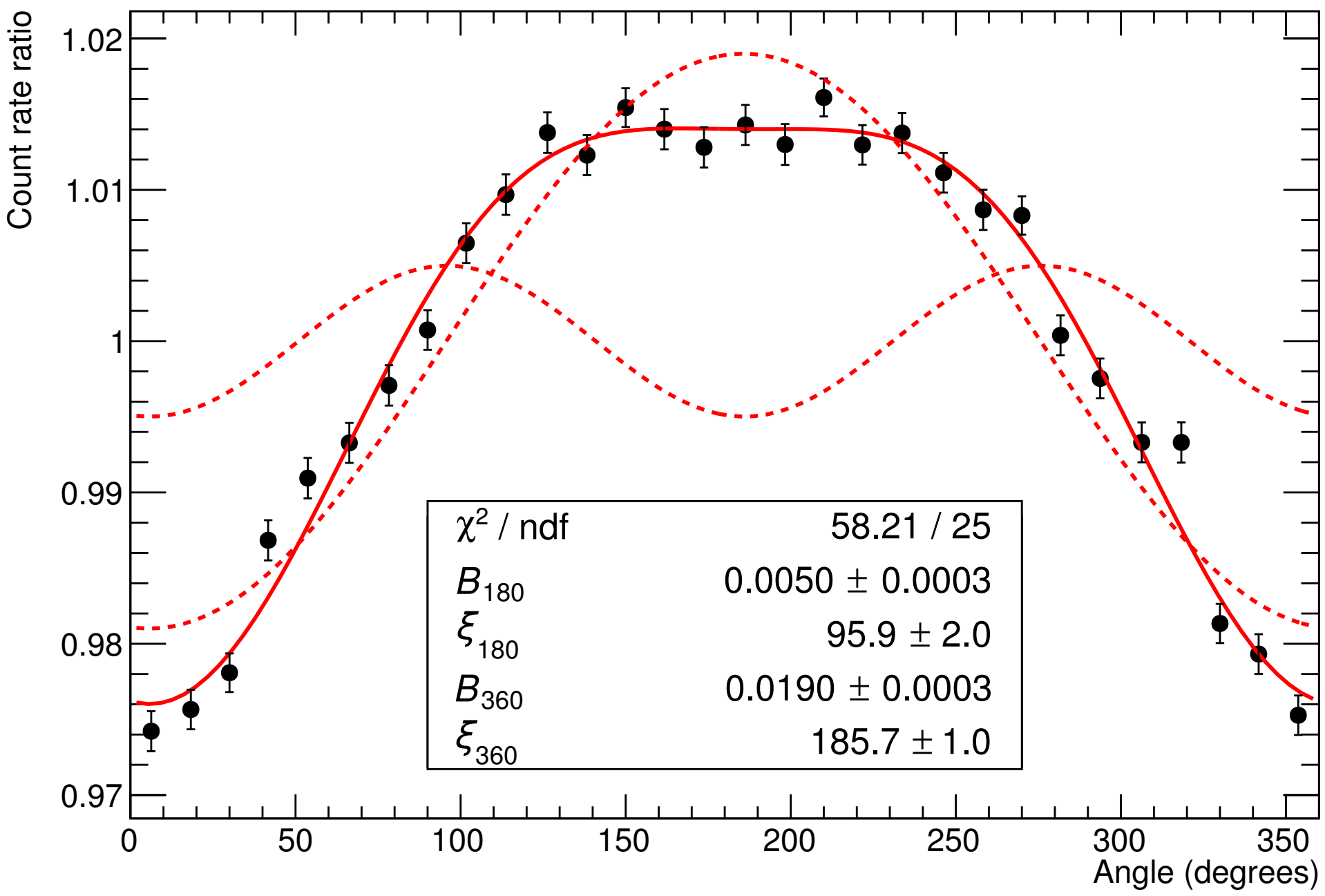

**Fig. S9. Relative count rates in SAS elements around the detector circumference.** Data points correspond to the 30 elements of the side anticoincidence shield, where each point has contributions from six detector units, separated by 60° relative to one another and de-convolved for the instrument rotation. Results are plotted as a function of position angle around the circumference of the detector, with 0° corresponding to zenith while 180° is the nadir direction. The solid line shows a modulation curve with both 180° and 360° components fitted to the data, while dashed lines indicate the individual components.

**Table 1. PoGO+ energy range and $M_{100}$.**

Results are shown for phase-integrated, nebula only and pulsar only data-sets. The variations in energy range reflect the difference in the spectral index between the Crab nebula and the pulsar.

| Phase selections | Energy range (keV) | $M_{100}$ (%) |
|---|---|---|
| Phase-integrated | 18 – 159 | 42.93 ± 0.74 |
| Nebula only | 18 – 155 | 42.82 ± 0.74 |
| Pulsar only | 18 – 176 | 43.41 ± 0.76 |

**Table 2. Polarization results for the Crab observation.**

Phase cuts as defined in Fig. 1 have been used. The maximum a posteriori probability estimate is given by $(p, \psi)$ and $(\sigma_p, \sigma_\psi)$. Since the measurement of the polarization fraction is positive-definite, $p > \Pi$ for all entries, i.e. maximum a posteriori result is always greater than the marginalized estimate of the polarization fraction. The angle is not governed by the same statistics and thus $\psi = \Psi$. Their uncertainties, however, are different, with $\sigma_\Psi > \sigma_\psi$ arising as a direct consequence of $p > \Pi$.

| Phase | MAP fraction $p$ (%) | Marg. fraction $\Pi$ (%) | 99% Upper Limit $\Pi$ (%) | MAP angle $\psi$ (°) | Marg. angle $\Psi$ (°) |
|---|---|---|---|---|---|
| All | 21.5 ± 4.9 | 20.9 ± 5.0 | 32.5 | 131.3 ± 6.6 | 131.3 ± 6.8 |
| OP | 19.5 ± 8.3 | $17.4^{+8.6}_{-9.3}$ | 37.3 | 137 ± 12 | 137 ± 15 |
| P1 | 20.0 ± 24.8 | $0^{+29}_{-0}$ | 72.6 | 153 ± 36 | 153 ± 43 |
| P2 | 39.9 ± 19.5 | $33.5^{+18.6}_{-22.3}$ | 80.9 | 86 ± 14 | 86 ± 18 |

**Table 3. Background anisotropy phase fits from polarization data and intensity maps.**

Comparison of background anisotropy phase fits (180° and 360° component) from SAS intensity maps, neutron data (LiCAF scintillators) and polarization data (SDCs). The systematic error in the phase angles from the intensity maps (first four rows) is estimated to be ±5°, since the start of the measurements is not exactly synchronised with the start of the polarimeter rotation, and since the pulse-height distribution data read-out every full minute results in about five seconds of dead-time per measurement which corresponds to 5° of rotation. In comparison, the statistical uncertainty is less than 0.002°. For the neutron detectors, the systematic error is estimated to be 10°, resulting from the fact that these detectors are mounted with elastic spacers inside the polarimeter and not fixed to the same mechanical structures inside the detector array. The larger deviation for the low-energy events reflects the fact that these are less penetrating and thus more affected by gondola structures surrounding the polarimeter. For the polarization data, field-rotation corrections are not applied, i.e. directions are considered relative to the gondola orientation, not relative to celestial objects.

| Data-set | 180° phase (°) | 360° phase (°) |
|---|---|---|
| SAS intensity map on-source low energy | 96 ± 5 | 385 ± 5 |
| SAS intensity map on-source high energy | 92 ± 5 | 369 ± 5 |
| SAS intensity map off-source low energy | 94 ± 5 | 374± 5 |
| SAS intensity map off-source high energy | 92 ± 5 | 369 ± 5 |
| LiCAF neutron detectors (neutron capture) | 108 ± 10 | 371 ± 10 |
| SDC waveforms off-source (no field-rotation) | 96.48 ± 3.49 | 363.3 ± 6.4 |